\pdfoutput=1 
\documentclass[sn-mathphys-num]{sn-jnl}% Math and Physical Sciences Numbered Reference Style 
\usepackage{graphicx}%
\usepackage{multirow}%
\usepackage{amsmath,amssymb,amsfonts}%
\usepackage{amsthm}%
\usepackage{mathrsfs}%
\usepackage[title]{appendix}%
\usepackage{xcolor}%
\usepackage{textcomp}%
\usepackage{manyfoot}%
\usepackage{booktabs}%
\usepackage{algorithm}%
\usepackage{algorithmicx}%
\usepackage{algpseudocode}%
\usepackage{listings}%
\usepackage{svg}
\usepackage[acronym, nogroupskip]{glossaries}
\usepackage{tabularx}
\theoremstyle{thmstyleone}%
\theoremstyle{thmstyletwo}%

\theoremstyle{thmstylethree}%

\makeglossaries
\newacronym{res}{RES}{renewable energy source}
\newacronym{rm}{RM}{reference model}
\newacronym{ncs}{NCS}{networked control system}
\newacronym{bess}{BESS}{battery energy storage system}
\newacronym{ccs}{CCS}{cloud control system}
\newacronym{lfc}{LFC}{load frequency control}
\newacronym{fm}{FM}{frequency meter}
\newacronym{bms}{BMS}{battery management system}
\newacronym{soc}{SoC}{state of charge}
\newacronym{tso}{TSO}{transmission system operator}
\newacronym[longplural={balance responsible parties}]{brp}{BRP}{balance responsible party}
\newacronym{bsp}{BSP}{balance service provider}
\newacronym{fcr}{FCR}{frequency containment reserve}
\newacronym{fcrn}{FCR-N}{frequency containment reserve normal}
\newacronym{fcrd}{FCR-D}{frequency containment reserve disturbance}
\newacronym{afrr}{aFRR}{automatic frequency restoration reserve}
\newacronym{mfrr}{mFRR}{manual frequency restoration reserve}
\newacronym{ffr}{FFR}{fast frequency reserve}
\newacronym{svk}{SvK}{Svenska Kraftnät}
\newacronym{ot}{OT}{operational technology}
\newacronym{dos}{DoS}{denial of service}
\newacronym{la}{LA}{load altering}
\newacronym{fdi}{FDI}{false data injection}
\newacronym{tds}{TDS}{time delay switch}
\newacronym{ev}{EV}{electric vehicle}
\newacronym{vpn}{VPN}{virtual private network}
\newacronym{dns}{DNS}{domain name system}
\newacronym{smca}{SMCA}{sliding mode control algorithm}
\newacronym{pid}{PID}{proportional-integral-derivative}
\newacronym{ra}{RA}{remote access}
\newacronym{ntp}{NTP}{network time protocol}
\newacronym{it}{IT}{information technology}
\newacronym{nids}{NIDS}{network intrusion detection system}
\newacronym{gui}{GUI}{graphical user interface}
\newacronym{plc}{PLC}{programmable logic controller}
\newacronym{hmi}{HMI}{human machine interface}
\newacronym{ldr}{LDR}{load disturbance rejection}
\newacronym{iot}{IoT}{internet of things}
\newacronym{scada}{SCADA}{Supervisory control and data acquisition}

\begin{document}

\title{From Balance to Breach: Cyber Threats to Battery Energy Storage Systems}

%Cloud-Controlled
%Cyberattacks and Impacts in Cloud-Controlled Battery Energy Storage Systems
%Evaluation of Cyberattacks on Cloud-Controlled Battery Energy Storage System
%
%%=============================================================%%
%% GivenName	-> \fnm{Joergen W.}
%% Particle	-> \spfx{van der} -> surname prefix
%% FamilyName	-> \sur{Ploeg}
%% Suffix	-> \sfx{IV}
%% \author*[1,2]{\fnm{Joergen W.} \spfx{van der} \sur{Ploeg} 
%%  \sfx{IV}}\email{iauthor@gmail.com}
%%=============================================================%%

\author[1]{\fnm{Frans} \sur{Öhrström}}%\email{iauthor@gmail.com}
\author[1]{\fnm{Joakim} \sur{Oscarsson}}%\email{iiauthor@gmail.com}
\author*[2]{\fnm{Zeeshan} \sur{Afzal}}\email{zeeshan.afzal@liu.se}
\author[1]{\fnm{János} \sur{Dani}}%\email{iiauthor@gmail.com}
\author[2]{\fnm{Mikael} \sur{Asplund}}%\email{iiauthor@gmail.com}

%\equalcont{These authors contributed equally to this work.}

\affil[1]{\orgdiv{Sectra Communications}, \country{Sweden}}
\affil[2]{\orgdiv{Linköping University}, \country{Sweden}}

%%==================================%%
%% Sample for unstructured abstract %%
%%==================================%%

\abstract{
Battery energy storage systems are an important part of modern power systems as a solution to maintain grid balance. However, such systems are often remotely managed using cloud-based control systems. This exposes them to cyberattacks that could result in catastrophic consequences for the electrical grid and the connected infrastructure. This paper takes a step towards advancing understanding of these systems and investigates the effects of cyberattacks targeting them. We propose a reference model for an electrical grid cloud-controlled load-balancing system connected to remote battery energy storage systems. The reference model is evaluated from a cybersecurity perspective by implementing and simulating various cyberattacks. The results reveal the system's attack surface and demonstrate the impact of cyberattacks that can criticaly threaten the security and stability of the electrical grid.}%including denial of service, false data injection, and load altering attacks affecting the grid balance.}


\maketitle

\section{Introduction}

\begin{figure}[h]svg
	\centering
	%\includesvg[width=1.0\textwidth]{figures/tug_of_war.svg}
	\includegraphics[width=1\textwidth]{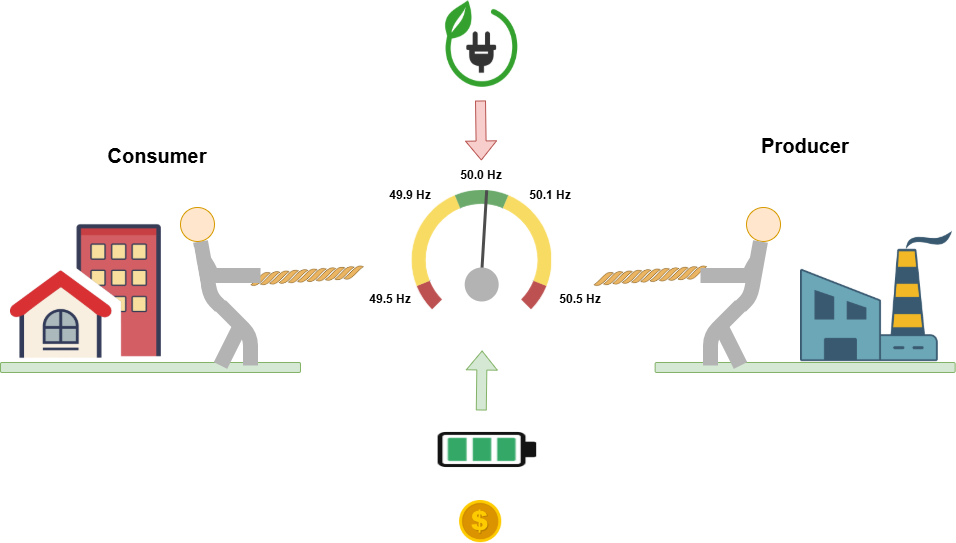}
	\caption{The balance between grid production and consumption. }
	\centering
	\label{fig:tug_of_war}
\end{figure}

The electrical grid is a critical infrastructure that forms the backbone of the modern society. To maintain stable grid operation, there must be an almost perfect balance between power production and consumption~\cite{Rinaldi_2022}. The consequences of large imbalances could be serious, such as blackouts in given areas or damage to connected equipment, which may result in physical hazards such as fires. The grid balance is estimated in real-time by measuring the alternating current (AC) frequency of the grid. If the frequency is 50 Hz, the consumers and producers are in balance. As shown in Figure \ref{fig:tug_of_war}, increased power demand from consumers can drag the frequency down if the production cannot keep up, while reduced power demand causes the frequency to rise if production does not decrease.

Coincidentally, there is also a constant and growing demand for electricity around the world. This creates a strong incentive to expand electricity production and meet the growing need. Moreover, the global transition towards a sustainable a zero-emission energy system is increasingly resulting in integration of distributed renewable energy sources (RESs) such as solar panels and wind turbines to the grid. As the grid increasingly relies on RESs, maintaining grid balance and controlling the frequency given different loads is becoming a challenge. This balancing process is also known as load frequency control (LFC). Compared to the reliable but fossil fuel-based traditional resources, the power produced by RESs tends to vary substantially over time because of the weather~\cite{Oshnoei_2020, Saini_2023, Trevizan_2022, mohan2020comprehensive, Ibraheem_2023}. As a result, this uncertainty makes it difficult to accurately predict how much energy to generate and to adjust production in a timely manner.

One solution to counteract the balancing difficulties in the grid is to employ electricity buffers in the form of battery energy storage systems (BESSs). As the name suggests, BESSs allow storage of energy using batteries enabling actors to sell balancing ability for profit, which helps keep the grid stable in the event of unforeseen supply and demand. Such a frequency balancing system often consists of a BESSs remotely controlled by a cloud control system (CCS) to act as an energy buffer and maintain the balance between generation and consumption of electricity. This set-up depends on time-critical control loops over the internet that are responsible for the stability and continued function of the electrical grid. The CCS receives a state of the grid in the form of regular frequency measurements from a remote sensor, called a frequency meter (FM). It continuously decides if a remote BESS should absorb energy (charge) or release it (discharge) to help maintain the frequency near 50 Hz. 

The benefits of using a cloud instead of local control include advanced and resilient control algorithms, dynamic contracts, and higher-level planning~\cite{cloud2}. However, the cloud poses significant requirements on the security of network connections~\cite{cloud1}, making it very important to develop robust solutions that can withstand failures and threats and continue to perform their function. These cloud solutions expose the electrical grid to the Internet, enabling serious cybersecurity vulnerabilities. For example, attackers could exploit these vulnerabilities to disrupt site-to-site communication or inject false data into the network~\cite{Rinaldi_2022}. The most likely source of such attacks is a nation-state actor aiming to destabilize or damage another country’s electrical grid for political or military objectives. Alternatively, the threat could come from hacker groups seeking to exploit balancing systems for financial gain, either by demanding large payments through ransomware attacks or manipulating electricity prices to their advantage. Another possibility involves terrorist groups that intend to inflict widespread damage, targeting electricity-dependent industries and societies by causing blackouts and destroying critical equipment~\cite{powergridresiliance}.

Regardless of who carries out these cyberattacks or what their motivations may be, it is crucial to understand how BESS-based grid-balancing systems are interconnected and to analyse the potential impact of cyberattacks on them and their capability to maintain grid balance. This paper advances knowledge in this area by contributing a reference model (RM) of a CCS connected to a remote BESS and a FM for providing LFC to an electrical grid. Furthermore, it contributes a simulation-based evaluation of potential cyberattacks that could threaten the system modelled by the reference model. Specifically, the paper aims to address the following research questions:

%\begin{itemize}
\begin{enumerate}

	\item How are frequency balancing systems based on cloud-controlled battery energy storage systems structured and integrated within the power grid?
	
	%How are load frequency control systems structured and integrated within a power grid to ensure stable operation?

	%\item What are the most common cyberattacks that can destabilize the grid balance? 
	\item Which specific attack vectors can target frequency balancing systems, and what is the impact of attacks on grid frequency regulation and stability? %and potentially lead to grid destabilization?%, with respect to frequency control?
	
	%\item What are the potential effects and impacts of these attacks on the grid frequency?%and, what measures can be taken to mitigate them?
	%\item How do cyberattacks on frequency balancing systems impact grid frequency regulation and stability?
\end{enumerate}

%\end{itemize}

The rest of the paper is organized as follows. In Section~\ref{sec:related_work}, a literature study is conducted to explore the related work and the structure of different systems within the power grid. Next, the gathered knowledge is applied to design a representative RM of a system where a BESS, controlled by a CCS, is used to balance power consumption and production. This is detailed in Section~\ref{sec:ref_model}. Based on the RM, potential attack vectors and relevant cyberattacks along with techniques are identified. These attacks are then evaluated in Section~\ref{sec:evaluation} using a simulator developed to represent a simplified version of the RM system. Section~\ref{sec:discussion} provides a discussion on the results, while Section~\ref{sec:conclusions} concludes the paper. An extended version of this work is available at~\cite{thesis}.

\section{Related Work}
\label{sec:related_work}

%This section presents research related to this paper with the aim to put the presented work in a wider context. 

Cloud services that control operational technology (OT) systems have been studied extensively in the literature. Baumgart et al.~\cite{Baumgart_2019} study BESS that are connected to the manufacturer's servers, facilitating monitoring capabilities, and business models where the batteries are used for LFC. Similarly, Naseri et al.~\cite{Naseri_2023} describe a solution where battery management systems (BMSs) are connected to a cloud to build a digital twin of the physical batteries, easing their management by visualizing parameters such as state of health and using machine learning to decide when to charge or discharge the batteries. The cloud makes it possible to run control algorithms that are far more computationally intensive than what is possible on the BMS themselves, thus predicting use and charging patterns to keep the batteries in decent shape for longer~\cite{Kharlamova_2020}. It has also been shown that using a CCS to control BESSs, lowers installation and maintenance costs and increases system flexibility compared to traditional control systems~\cite{sanchez2019bibliographical}.

However, connecting OT systems, such as BESSs, to a CCS requires data to be sent over the internet, making them vulnerable to cyberattacks as demonstrated by~\cite{Naseri_2023, Kharlamova_2020}. If a BESS within the power grid is controlled via a cloud-based system, the communication pathways become potential targets for cyberattacks. These threats could compromise the link between the sensors and the cloud, as well as the connection between the cloud and the battery units~\cite{Kharlamova_2020}. Several literature reviews~\cite{mohan2020comprehensive,mohammad2023} in the domain of cyber-physical systems have revealed common attacks such as  time delay switch (TDS) attacks, false data injection (FDI) attacks, denial of service (DoS), replay attacks, and load altering (LA) attacks. Mahmoud et al.~\cite{mahmoud2019modeling} showed that distributed denial of service (DoS) attacks can cause instability of power grids and produce lengthy delays between packets being sent and received in networked control systems (NCSs). LA attacks can either be achieved by an attacker that physically controls a large enough part of the load (power) in an area or by somehow manipulating aggregated load communication of demand aggregation systems~\cite{mohan2020comprehensive}. Such aggregated attacks against the power grid are considered among the most critical threats~\cite{Trevizan_2022}. If an attacker gets access to enough load-affecting devices, such as electric vehicle (EV) chargers, they could coordinate consumption patterns that might impact the stability of the grid frequency. Another realistic scenario is presented in~\cite{Baumgart_2019} where servers hosted by BESS vendors can be controlled by an attacker to decide when batteries should push or pull energy from the power grid through control signals. Furthermore, frequency fluctuations (whether due to load changes, faults, or cyberattacks) in a local area can leak into other areas via tie-lines, and in the worst case threaten the stability of the overall network or even cause a blackout of the entire power grid if the frequency deviance is large enough~\cite{mohan2020comprehensive, Wu_2019}. Moreover, OT systems are often monitored and controlled remotely using solutions such as a virtual private network (VPN)~\cite{Trevizan_2022}. This could enable an attacker to find a security hole and gain access to the network via the VPN connection. Another type of common vulnerability is poor encryption in the communication between BESSs installed in homes and the manufacturer servers used for remote control~\cite{Baumgart_2019}. Yet another concern is CCS's reliance on external services, such as domain name system (DNS) servers, which could expose the system to DNS hijacking attacks. In such attacks, an attacker redirects traffic to a server under their control, as highlighted by~\cite{Chauhan_2023}.

%Based on this review, we shortlist the following attacks to be evaluated as part of this work.

On the defensive side, different approaches can be used to detect and mitigate attacks in cyber-physical systems. Some monitoring-based approaches compare the current state of the system with a model of how it should look like, called a digital twin. If they detect a significant difference, the system knows it is under attack or operating under abnormal conditions~\cite{Wu_2019, Akbarian_2023}. Attacks such as LA attacks can be mitigated by a  sliding mode control algorithm (SMCA) together with a BESS~\cite{Rinaldi_2022} and false data injection (FDI) attacks can be mitigated by making use of signal processing and blockchain~\cite{Ghiasi_2021, Trevizan_2022, mohammad2023}. Gumrukcu et al.~\cite{Gumrukcu_2022} present another form of FDI attack called hijacking, where the control signals are not slightly altered but completely forged by the attacker. They propose a solution where distributed screening is used to identify attacks and a fault detection metric is used to distinguish between attacks and sensor faults.

Simulations are commonly used in the field of power systems to evaluate attacks against power systems and assess defence strategies. Different simulators have been proposed and developed in the literature for this purpose. One example is GridAttackSim~\cite{GridAttackSim}, which is a smart grid attack simulation tool that combines the NS-3 and GridLAB-D simulators with co-simulator FNCS to make a realistic representation of cyberattacks on an electrical grid~\cite{GridAttackSim}. Cosima~\cite{cosima} is another attempt at creating a realistic yet performant smart grid simulator~\cite{cosima}. It combines OMNeT++ which is a C++ framework typically used for computer network simulation, with the co-simulation framework Mosaik\footnote{\url{https://mosaik.offis.de/}}. Kumar et al.~\cite{Kumar_2016} created a stand-alone microgrid simulation consisting of several RESs and loads. They used a proportional-integral-derivative (PID) controller to balance the system frequency and showed that a load disturbance rejection (LDR-tuned) PID reduces transient periods from that of a classical PID. 

Despite extensive research on cloud-based control systems for OT systems and BESSs, gaps remain in understanding how these systems integrate within power grids to balance the power consumption and production. Specifically, there is limited work on developing a comprehensive model and studying the operational integration of CCSs within the grid. The complex interactions between cloud infrastructures, communication pathways, and power grid components need further exploration to develop a reference model. Additionally, while cyberattacks on power grids have been widely studied, few focus on the unique vulnerabilities of cloud-based frequency control systems. Moreover, the impacts of cyberattacks on grid frequency regulation and stability have not been thoroughly examined. There is also a lack of comprehensive simulation models that realistically capture the dynamics of cloud-controlled BESSs under attack, particularly in the context of real-time frequency regulation. This paper addresses these critical gaps by investigating the attack vectors targeting frequency-balancing cloud control systems and evaluating their impact on grid stability. By doing so, the work aims to enhance the security of cloud-integrated frequency control systems within power grids, ensuring their safe and reliable operation in the face of evolving cyber threats.

\section{Reference Model}
\label{sec:ref_model}

\begin{figure}[h]
	\centering
	\def\svgwidth{\columnwidth}
	\includegraphics[width=1.1\textwidth]{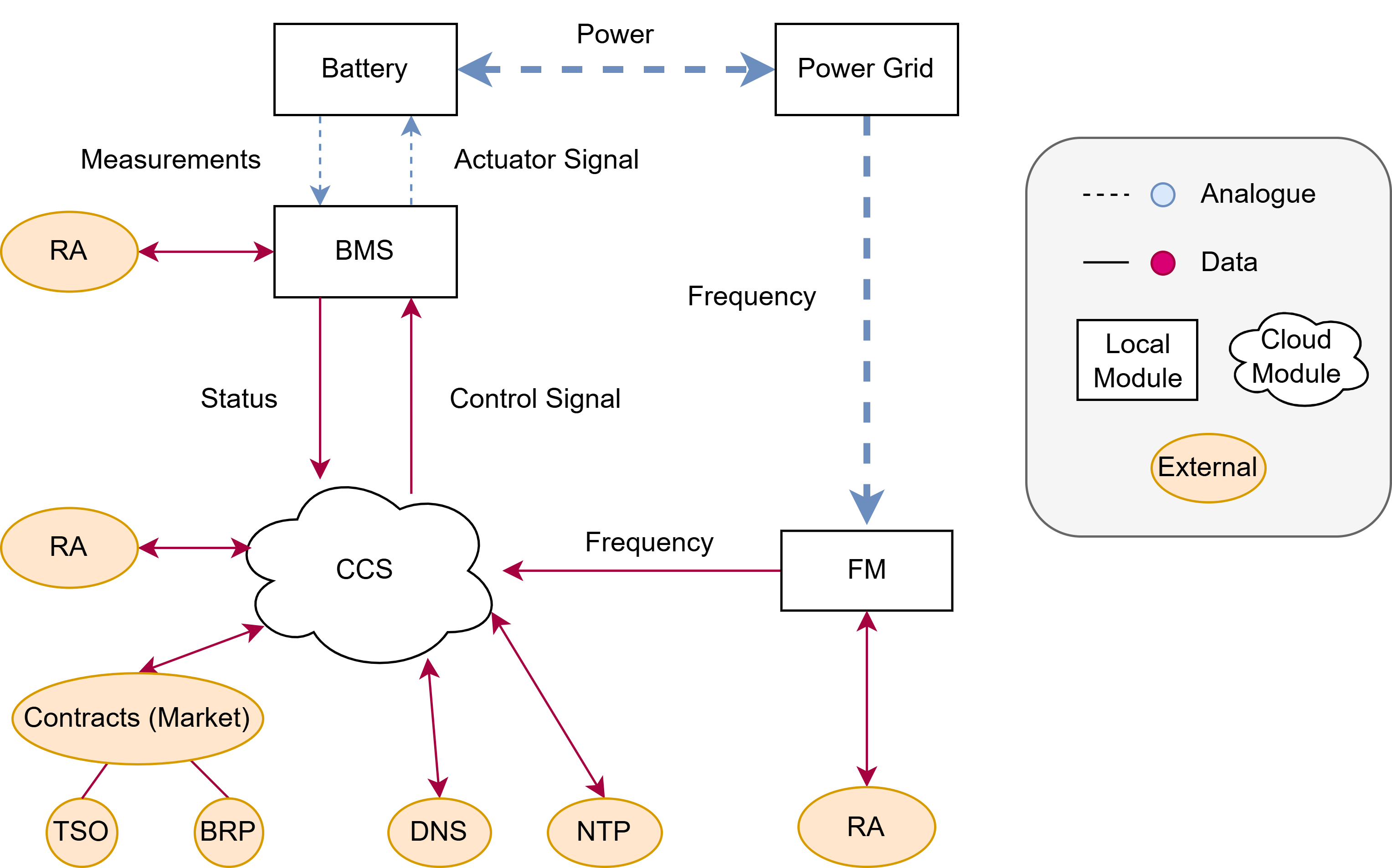}
	\caption{The proposed reference model.}
	\label{fig:reference_model_grid}
\end{figure}

This section proposes a reference model for cloud-controlled BESS. Based on this RM, various attack vectors and potential cyberattacks along with countermeasures are identified in the upcoming sections. The RM was developed through a highly iterative process, which involved an investigation of relevant literature and  iterative feedback from domain experts within the CyREC project~\cite{Cyrec}. The analysed literature consists of peer-reviewed articles addressing the structure of electrical grids, from distribution-level end-points to transmission systems, as well as energy storage systems, control systems, and mechanisms for balancing energy production and consumption. Additionally, publications on common cybersecurity issues and attack vectors in similar critical infrastructure systems were analysed. The review also identified key actors and stakeholders involved in the balancing market. Insights from this review informed the initial design of the RM, which was subsequently presented during regular project meetings. In the meetings, domain experts, including a grid operator and an aggregator, cybersecurity engineers, and researchers, provided structured feedback on its components and alignment with real-world scenarios. The RM was refined iteratively based on this feedback, with each updated version discussed and evaluated in subsequent meetings to ensure relevance and practical applicability. Figure \ref{fig:reference_model_grid} illustrates the proposed RM, encompassing the key modules, components, links, and communication flows integral to the system. The model is designed to encapsulate the critical elements and their interactions, providing a structured framework for understanding the system's architecture and dynamics. To ensure consistent interpretation, the following definitions for the RM elements are adopted and will be referenced throughout the remainder of the paper.

\begin{itemize}
	\item Module: an independent component within the closed-loop system%, excluding internal or external components.
	\item Link: either an analogue or data connection between two components.
	\item Component: any part of the RM that is not a link. This includes modules or parts that are either internal to a module or external to the closed-loop system.
	%\item Communication: any flow of data.   
	
\end{itemize}

\subsection{Modules}
The proposed RM in Figure \ref{fig:reference_model_grid} contains the following modules: a generalization of the power grid, one or more FMs, a CCS, and one or more BMSs, each connected to one or more batteries. These modules are described below.

%\paragraph{Power grid}
\subsubsection*{Power grid}
The power grid module is an abstraction of a distribution-level electricity grid in the area that needs balancing of the alternating current frequency. The balance is affected by factors such as unpredictable energy consumption by households or industries, fluctuating output from RESs, and actions of other generation actors such as nuclear and fossil fuel power plants trying to meet the demand.

\subsubsection*{Frequency meter (FM)} 
AC frequency is periodically measured in a real power grid at multiple locations (from distribution side all the way closer to the transmission) using devices such as phasor measurement units (PMUs). In the proposed RM, these devices are represented by the FM which regularly (e.g. once every second) reads the power grid frequency and forwards it as a network packet over the internet to the CCS. 

\subsubsection*{Cloud-control system (CCS)} This is the central hub of the control loop, where measurements from one or more FMs are received and fed to a control algorithm that produces output in the form of control signals sent to BMSs. The control signals tell a BMS how much power its batteries should push or pull from the power grid to keep the frequency stable. The CCS may have to coordinate many BMSs, thus it has the responsibility for spreading out the load on these different BMSs, both for increasing the longevity of their batteries and for the average state of charge (SoC) in the batteries to stay close to around 50-60\% to keep them flexible and ready for more balancing. Given that cloud systems typically offer greater computational power than conventional on-site systems, and can coordinate many FMs and BMSs, the CCS enables resilient control algorithms, cyberattack detection, and other advanced security measures. These capabilities are further detailed in Section \ref{final_RM_servies}.

\subsubsection*{Battery management system (BMS)} The BMS controls the batteries on site and houses the intelligence needed to handle them correctly, for example by ensuring that batteries operate within a bounded SoC range and are not overheated or otherwise put in an unhealthy operational state. The BMS receives control signals from the CCS and actuates them using its batteries. The BMS also reads measurements such as SoC from each battery and sends status messages back to the CCS, which can help the CCS coordinate multiple BMSs efficiently.

\subsubsection*{Battery} A battery is a hardware unit that can push or pull electricity to and from the grid. Depending on the market in which the CCS operates, different requirements on the battery are imposed, such as the maximum power it should push or pull, and the duration for which it must maintain balance in a specific direction without becoming empty or full. 

\subsection{External components}

The RM model in Figure \ref{fig:reference_model_grid} also contains components or infrastructure external to the closed-loop system. These components include DNS and network time protocol (NTP) servers, terminals for remote access (RA) from the outside, and contracts with other actors such as a transmission system operator (TSO) and balance responsible parties (BRPs).

\subsubsection*{Remote access (RA)} As it is convenient for the users and manufacturers of a BESS to monitor and maintain the systems remotely, many manufacturers offer this service as a cloud solution~\cite{Baumgart_2019}. Therefore, the RM contains three RA points. One RA point is connected to the CCS, where system administrators may want to affect the control strategy manually, update software, and make changes to and/or monitor nodes and network traffic in the CCS from the outside. The other two RA points, connected to the BMS and FM, could also benefit system administrators and operators by enabling remote monitoring and firmware upgrades. The RA links could additionally open up the possibility of outsourcing cybersecurity and monitoring to a security operations center, overseeing hosts and network traffic to detect and stop cyberattacks.

\subsubsection*{Contracts (TSO and BRP)} The CCS operates in a balancing flexibility market. The market dictates how much power the batteries, controlled by a CCS, should be able to push/pull and for how long. The contracts component represents this information.

\subsection{Communication links}

The following section describes the two communication link types defined in the RM.

\subsubsection*{Analogue signals} The dotted bold blue lines in Figure \ref{fig:reference_model_grid} represent continuous analogue signals, e.g. power flow between a battery and the power grid, the grid frequency measured by analogue sensors in the FM, and signals sent to and from batteries by their management systems. These connections are physical, usually within short distances, and are not part of the cyber realm or connected to the internet. As a result, while the connections could be sabotaged by cutting cables, such attacks do not fall under cyberattacks and are therefore beyond the scope of this paper.

\subsubsection*{Data Connections} The CCS is connected using some public network such as the internet not only to the FM and BMS but also to external infrastructure such as a DNS and NTP server, RA terminals, along with the trading system used to sign contracts with other parties (e.g. TSOs and BRPs). These publicly accessible connections are possible attack vectors that must be carefully examined before deployment in a real system.

To ensure data integrity and maintain the performance of real-time communication between the FM and CCS, as well as between the CCS and BMS, a secure transmission protocol (such as the WebSocket protocol~\cite{Fette_2011}) is used to encode messages as they are sent over the internet. This protocol should authenticate and timestamp packets for integrity assurance, and could also benefit from being encrypted to combat sniffing of the data (which could be used as part of a cyberattack). Integrity is vital: tampered values could be detrimental since they might cause the CCS to make the wrong decisions, possibly leading to destabilized batteries or even harming the power grid instead of providing balancing.

In addition to the secure transmission protocol inside the control loop, we include some other protocols that a CCS might use. The DNS protocol is used for the cloud to reach and be reached by other servers using domain names. NTP is used for time synchronization, which is then used for timestamping packets. The RM also includes some RA links to the CCS, BMS, and FM, which could utilize different VPN, remote desktop, or remote shell protocols (such as OpenVPN, IPSec, RDP, VNC, SSH, etc.). These external links reduce introduce additional attack vectors, particularly as they may be misconfigured or poorly maintained. However, these protocols are not investigated in the paper.

%aspects
\subsection{Services of the CCS}\label{final_RM_servies}

\begin{figure}[h]
	\centering
	\def\svgwidth{\columnwidth}
	%\includesvg{figures/CM_4.6.svg}
	\includegraphics[width=1\textwidth]{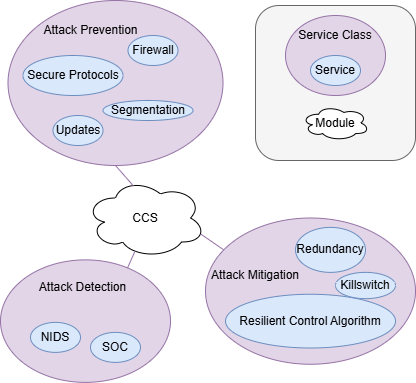}
	\caption{Services provided by the CCS.}
	\label{fig:reference_model_cloud}
\end{figure}

We also consider some additional security services in the reference model provided by the CCS, aimed at enhancing the security and robustness of the system. These services are crucial for maintaining grid frequency stability despite the presence of cyberattacks. Figure \ref{fig:reference_model_cloud} illustrates these services, with a description provided below. 

\subsubsection*{Attack prevention} Designing a robust and secure CCS that by design helps prevent cyberattacks from occurring is as important as implementing effective detection and mitigation mechanisms. This service highlights information technology (IT) standard security practices, including the use of secure communication protocols, regular patching of systems with the latest software updates, and the implementation of segmented networks with properly configured firewalls to prevent attackers from gaining access to the most critical assets.

\subsubsection*{Attack detection} As the RM system operates as a time-sensitive control system dependent on network links, it is crucial that cyberattacks can be quickly detected. The control system may not function optimally if there are losses or delays in frequency measurements or control signals. Integrity attacks, such as an FDI or replay attack, could modify the frequency and control signal values. If undetected, these attacks could cause the controller or BMS to trust manipulated data, potentially resulting in severe consequences such as financial losses or even catastrophic events.

There are many ways to detect cyberattacks. For the RM, we focus on a network intrusion detection system (NIDS) that autonomously scans and detects strange behaviours in a network and a security operations center that in addition to autonomous systems has human resources to monitor network traffic and systems and detect anomalies. Using a NIDS and maintaining a security operations center to analyses traffic captured by logging systems are well-established industry practices for defending against cyberattacks~\cite{Behera_2022, Ganesan_2017}. These methods do not constitute a complete list, but rather represent conceptual solutions exemplifying solutions that can be implemented.

\subsubsection*{Attack mitigation} During a cyberattack, immediate actions must be taken to mitigate its impacts. Without such mitigation, the grid stability could be compromised as a result of significant frequency fluctuations, potentially resulting in power blackouts or damage to connected equipment. An important mitigation is a resilient control algorithm that can continue operating the system despite an ongoing attack~\cite{mohan2020comprehensive}. Another defensive technique is redundancy, such as deploying multiple FMs and BMSs. Thus, if one module or link is compromised, the system can disregard it. Additionally, a kill switch ability could be implemented, allowing the system to enter a fail-safe mode and continue to operate even if the CCS is compromised.

\section{Attack Impact Evaluation}
\label{sec:evaluation}

This section provides an evaluation of various cyberattacks. It starts by introducing the simulator used to instantiate the reference model and evaluate the cyberattacks. Next, a detailed description of the attacks is provided, and their impacts are examined.

%The final simulator architecture can be seen in Figure \ref{fig:simulator_model}. 

%\begin{figure}[h]
%    \centering
%   \def\svgwidth{\columnwidth}
%    \includesvg{figures/SIM/SIv4.svg}
%    \caption{Diagram of the final simulator architecture, used for the evaluation (component and link types are described in the grey box).}
%    \label{fig:simulator_model}
%\end{figure}

\subsection{Simulator}

%This is a standard practice in the field due to the high risks of testing attacks on live systems.
%Additionally, many of these simulators abstract key aspects of cyberattacks, such as the precise timing, propagation, and impact of malicious actions on grid components. Such oversimplifications make it difficult to assess the nuanced effects of attacks on grid stability and performance, which are central to this study. Moreover, these simulators often fail to account for specific grid control strategies, cybersecurity measures, or attack vectors that are critical for the evaluation of real-world cyber threats.

\begin{figure}[h]
	\centering
	\def\svgwidth{\columnwidth}
	\includegraphics[width=1\textwidth]{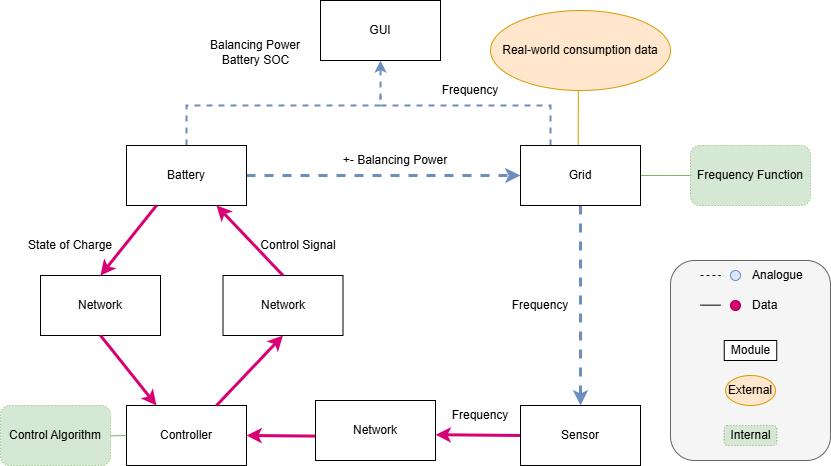}
	%\includesvg{figures/SIM/SIv4.svg}
	\caption{Simulator architecture.}
	\label{fig:simulator_model}
\end{figure}

We employ simulations to evaluate a set of relevant cyberattacks. While several simulators are available in the literature, as discussed in Section~\ref{sec:related_work}, they present limitations that hinder their use for our goal. These existing tools are often outdated, rely on legacy framework, or are designed with higher-level goals. More specifically, existing simulators primarily focus on general-purpose grid models, which often fail to capture the unique characteristics of cloud-controlled BESS in grid operation. As a result, they do not offer the granularity required to model and evaluate the precise impact of cyberattacks targeting BESSs. Adapting these tools to meet the precise requirements of our attack scenarios could lead to inaccurate or incomplete representations of the system's behaviour under attack. Given these shortcomings, we chose to design and implement a custom simulator tailored to the unique objectives of this work. This approach enables a more detailed, accurate, and rigorous evaluation of cyberattacks, ensuring that the simulator reflects specific characteristics of grid systems under cyberattacks. While we use a custom simulator, the insights and results derived from this work are generally applicable. The custom simulator is inspired by existing simulators such as GridAttackSim~\cite{GridAttackSim}, Maestro~\cite{Maestro_2020}, and Cosima~\cite{cosima}, and uses co-simulators based on HELICS~\cite{HELICS}. The architecture of the simulator can be seen in Figure \ref{fig:simulator_model}. Additional details regarding the design choices and implementation details of the simulator fall beyond the scope of this paper, and the interested reader is referred to~\cite{thesis}. The simulator's source code, along with a user manual, is openly available at~\cite{repo}.

%This figure and the mention of it should probably be removed as we do not want to focus on the simulator

Once the simulator is in place, it is used to instantiate the RM. This enables assessment of how various cyberattacks could destabilize the system and provides insights into the potential impacts of cyberattacks on the real system it represents. The evaluation is based on a number of attacks as identified in Section~\ref{sec:related_work}. The simulated system's stability (robustness) during attacks using different configurations is evaluated by observing oscillations, deviance, and steady-state error of the simulated grid frequency. The frequency $f$ is calculated in Hz by the simulator using the following formula: 

\begin{equation}\label{frequency_consumption_relationship}
	f (Hz) = \frac{production}{consumption} \times 50 Hz
\end{equation}

This relationship is a simplified representation of reality, where the frequency reduces linearly when consumption increases or production decreases, and vice versa~\cite{SVK_frekvensstabilitet, modig2022overview}. The criteria for an oscillating system is an alternating frequency in the live graph. Deviance indicates that the frequency value increases or decreases continuously until moving far from the nominal frequency (between 49.9 and 50.1 Hz). Finally, the requirement for steady-state error is that the frequency eventually stabilizes around a constant value outside the nominal frequency. 

The simulator includes many parameters. Some parameters are altered when conducting different cyberattacks, while others are kept static throughout the evaluation. These parameter choices are informed by discussions with domain experts. The default system, referring to the parameter configurations that remained unchanged, is defined in Table \ref{tab:default_simulator}.

\begin{table}[h!]
	\caption{The default simulator parameters.}
	\label{tab:default_simulator}
	\begin{tabular}{ll}
		\hline
		\textbf{Parameter}  & \textbf{Value}                           \\ \hline
		Battery power       & 2 MW                                     \\
		Battery capacity    & 2 MWh                                    \\
		Sensor interval     & 50 (send measurements once every second) \\
		%Production delay    & 0s (no production change)                \\
		Consumption scaling & 0.02                                     \\ \hline
	\end{tabular}
\end{table}

\subsection{Data and battery capacity}\label{evaluation_subsection_scaling_consumption}

\begin{figure}[H] 
	\centering 
	\includegraphics[width=1\textwidth]{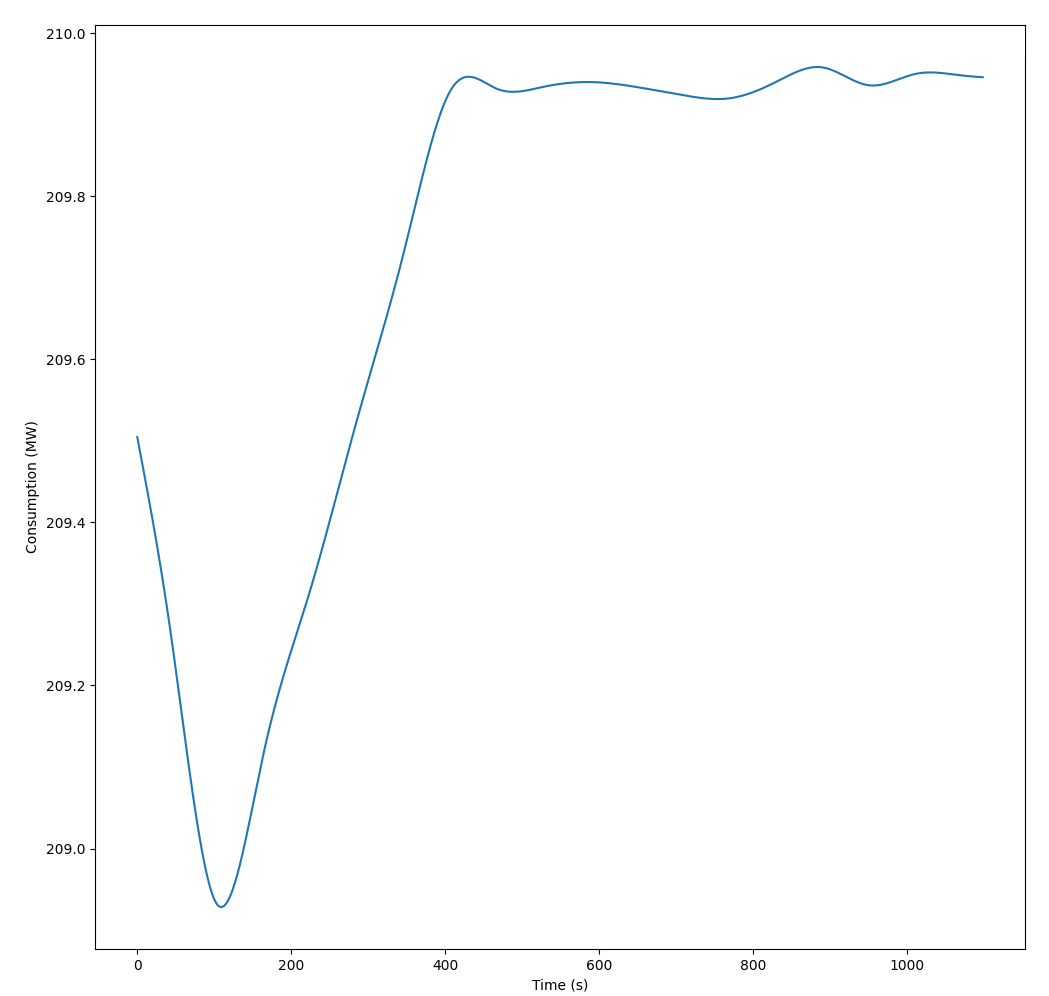}
	\caption{Scaled consumption data.}
	\centering
	\label{fig:ev_consumption}
\end{figure}

The simulator uses real world consumption data for evaluations. This data is visualized in Figure \ref{fig:ev_consumption} and provided by the SvK TSO agency in Sweden~\cite{SvK_kontrollrummet}. The data consists of minute-wise electricity consumption values in Sweden from the summer of 2023. From these, minutes 38 to 120 from the July 2023 dataset is used in the evaluations, but scaled down by a factor of 0.02 from the size of approximately 10,500 MW in the original dataset to about 210 MW.  This is estimated to be the consumption in a city with about 100,000 households, a reasonable size for a distribution grid that a threat actor could target.

The scaling of consumption data is performed as follows. An average Swedish household consumes approximately 20 MWh per year~\cite{Vattenfall_elkonsumption}. Dividing this number by the total number of hours in a year (8760) gives an average power consumption of 2.28 kW per household. This value is then multiplied by 100,000 to account for the number of households in the hypothetical city used for our evaluation, yielding a total power consumption of approximately 228 MW. This results in a scaling factor of approximately 0.0217, obtained by dividing 230 MW with 10,500 MW, which is the original power baseline. For simplicity, a consumption scaling factor of 0.02 is used, which corresponds to a theoretical city power consumption of approximately 210 MW. 

%The same scaling factor used to scale the power consumption is used to scale down the battery power from the current total estimate of 107 MW in Sweden~\cite{Ny_teknik}: 

%\begin{equation}\label{frequency_consumption_relationship_with_inertia}
%round\_to\_integer( 107 MW \cdot 0.02 ) = 2 MW
%\end{equation}

The same scaling factor of 0.02 is used to scale down the battery power from the current total estimate of 107 MW in Sweden~\cite{Ny_teknik}. Thus, the battery energy storage capacity, representing the maximum state of charge, is set at 2 MWh to match it with the power output of 2MW. This value is based on the specifications provided by one of the most widely used household batteries on the market~\cite{Pixii}.

\subsection{Evaluation}

In the following subsection, an evaluation of various cyberattacks and their impacts on the grid frequency is provided. 

\subsubsection{Time delay switch (TDS) attack}

\begin{figure}[H] 
	\centering 
	\includegraphics[width=1.1\textwidth]{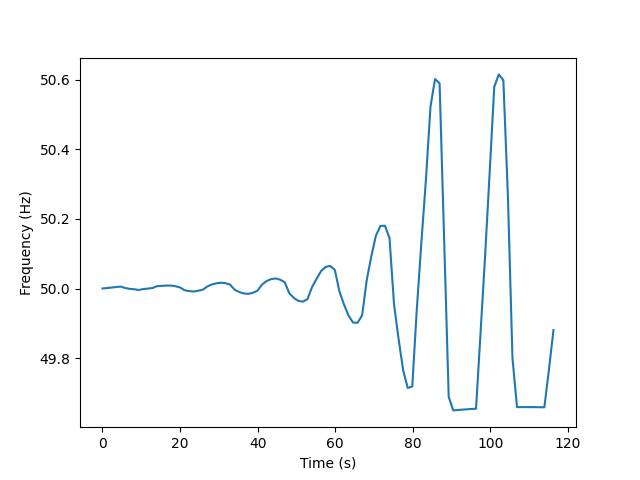}
	\caption{A TDS attack.}
	\centering
	\label{fig:ev_delay}
\end{figure}

An attacker can cause oscillations in the grid frequency by introducing a delay in the communication between the sensor and the controller or between the controller and the battery. Figure \ref{fig:ev_delay} shows an attack scenario where a delay of 4 seconds is introduced in the sensor-controller link causing an increasing oscillation in the signal.%, while the controller is configured to be more reactive than its default tuning. % (Kp=5, Ki=2, Kd=0). 

Experiments were also conducted by running the simulator with different delays on the links, observing that a random delay attack may be more difficult to detect and mitigate than a constant delay attack. Delays spread over a uniform distribution with a minimum value of 8 seconds and a maximum value of 12 seconds occasionally cause the grid frequency to oscillate, even though most delay values (on average) are lower than 11 seconds. At the same time, we discovered no meaningful difference between introducing the delay between the sensor and the controller or between the controller and the battery. This outcome is somewhat expected, as the system is a closed-loop feedback system, meaning the location of the delay should not matter as it will slow down the feedback regardless. Finally, we observe that a uniform distribution of delays with a minimum value of 0 seconds and a maximum value of 12 seconds does not destabilize the default system. This indicates that occasional lag spikes of high values do not lead to oscillations or instability.

\subsubsection{Denial of service (DoS) attack}

A DoS attack could target the network communication within the simulator, such as the pathways between the sensor and the controller or between the controller and the battery. In this context, a DoS is characterized by an attacker causing between 0\% and 100\% of the network packets on a link to be dropped. This results in reduced information available to the receiver. For example, if the controller experiences such an attack, it may receive incomplete measurement data, affecting its ability to generate accurate control signals for the battery. 

\begin{figure}[H] 
	\centering 
	\includegraphics[width=1\textwidth]{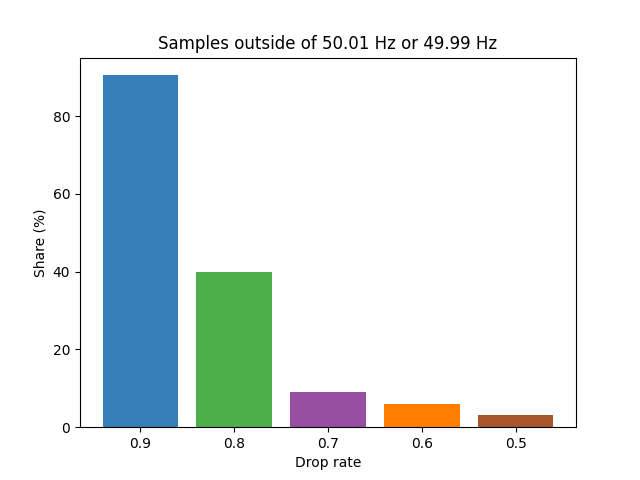}
	\caption{Comparison of drop rates showing the share of abnormal frequency values.}
	\centering
	\label{fig:ev_drop}
\end{figure}

A number of tests are run using the controller's default tuning and different DoS rates of dropped packets (between 0.0 and 1.0). Figure \ref{fig:ev_drop} shows that a drop rate from 0.8 (i.e. 80\% of packets dropped) and upward results in unpredictable simulator behaviour with many frequency measurements outside 50.01 Hz or 49.99 Hz. In contrast, drop rates of 0.7 and below have a much smaller impact, with only a few frequency measurements straying outside the 49.99 Hz to 50.01 Hz range, and none exceeding the normal operating boundaries (below 50.1 and above 49.9 Hz). These results suggest that the system we tested is relatively robust to packet loss. However, if the drop rate is 0.8 or higher, it would be wise to turn the controller off since it may do more harm than good. The tests also suggest that higher drop rates can be mitigated by reducing the sensor interval, effectively sending more frequency measurements per time unit to ensure the controller receives sufficient information despite packet loss. 

\subsubsection{False data injection (FDI) attack}

A FDI attack targets network communication by intercepting traffic and modifying the data exchanged between components, thus controlling what the receiver sees. For instance, such an attack can manipulate the frequency measurements from the sensor to the controller or the control and status signals transmitted between the controller and the battery. 

In the simulator, FDI attacks are defined by several parameters: interval, offset, randomness, base, and scaling factor. The interval dictates how often a packet is modified. An interval value of 1 signifies that every packet is modified, a value of 2 means every other packet, and so forth. The offset parameter adds or subtracts an offset to the original value, e.g., to the frequency measurements or control signals. Randomness adds a uniformly distributed randomness to the original value, e.g., to the frequency measurements or control signals, drawn from a specified range (upper and lower bounds). Base replaces the original value with a new constant value. Finally, the scaling factor multiples the original value by a specified scaling factor to change its magnitude (e.g., a scaling factor of 2 doubles the original value). To experiment with more sophisticated FDI attacks, we employ a script to dynamically adjust these parameters during simulator execution. This allows the attacks to adapt based on variables such as time, frequency, or balancing power. This approach surpasses manual parameter configuration via the initialization file or GUI, enabling more complex attack scenarios, as described in the following.

\paragraph{Ramp attack}

\begin{figure}[H] 
	\centering 
	\includegraphics[width=1.1\textwidth]{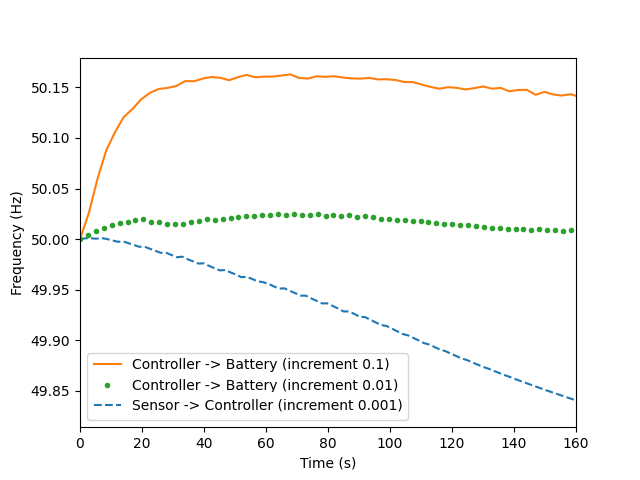}
	\caption{Three examples of ramp attacks.}
	\centering
	\label{fig:ev_ramp}
\end{figure}

A ramp attack gradually increases or decreases a value over time, making it difficult for the detection system to identify the attack. The simulations show that this type of attack is particularly effective when targeting the frequency measurements, as it makes the controller slowly lose track of the system state, resulting in frequency deviations. As shown in Figure \ref{fig:ev_ramp}, the sensor to controller line slowly reaches outside the normal operating boundaries, making the attack difficult to detect from the controller's perspective. The figure shows three examples of ramp attacks illustrated by different colours. In the first example, the sensor-controller link value is increased by 0.001 per second, causing an increasing error over time. Then, the controller-battery link value is incremented by 0.01 and 0.1 per second, causing the frequency to stabilize with small or large steady-state errors depending on the rate of the ramp.

Interestingly, targeting the control signal with a ramp attack has mixed results. The increment needs to be large enough for the frequency to deviate noticeably from its nominal value, eventually stabilizing at a steady-state error.

\begin{figure}[H] 
	\centering 
	\includegraphics[width=1.1\textwidth]{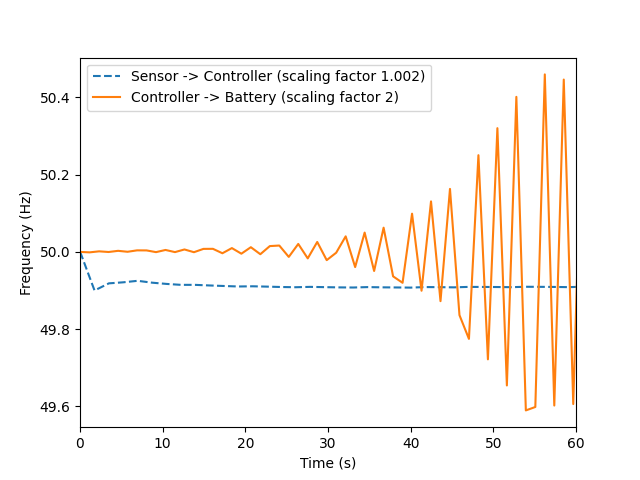}
	\caption{Scaling attacks on the frequency measurements.}
	\centering
	\label{fig:ev_scaling}
\end{figure}

%t adversaries will randomly inject reduced values
%into original data when launching scaling attacks, 
\paragraph{Scaling attack}
In a scaling attack, an adversary amplifies data by applying a scaling factor to the original value. This alters the data in a way that may not be immediately obvious. We experimented with different scaling factors and targeting different links and control signals. Simulation results show that the impact of such attacks on the system behaviour varies depending on the scaling factor and whether the sensor-controller measurements or controller-battery control signals are targeted. Figure \ref{fig:ev_scaling} shows the results when the first link (sensor to controller) is targeted. The system experiences a steady state error exactly equal to the difference between the scaled and unscaled values. For example, a scaling factor of $1.002$ causes a steady state error of 0.1 Hz and the system stabilizes at around 49.9 Hz as it believes it is operating at $50 \times 1.002 = 50.1$ Hz instead of 50 Hz. Whereas, a scaling attack on the control signals (controller to battery) can have a bigger impact depending on the scaling factor. A scaling factor of $1.1$ does not cause any noticeable difference however, a scaling factor of 2 causes the system to oscillate its frequency with an increasing amplitude over time (as shown in~Figure \ref{fig:ev_scaling}). Thus, we conclude that, if the scaling factor is sufficiently large, this overreaction can drive the system into complete instability. 

%This occurs because the controller overcompensates for the artificially amplified signals, creating a feedback loop that amplifies deviations.

\paragraph{Pulse attack}

\begin{figure}[H] 
	\centering 
	\includegraphics[width=1.1\textwidth]{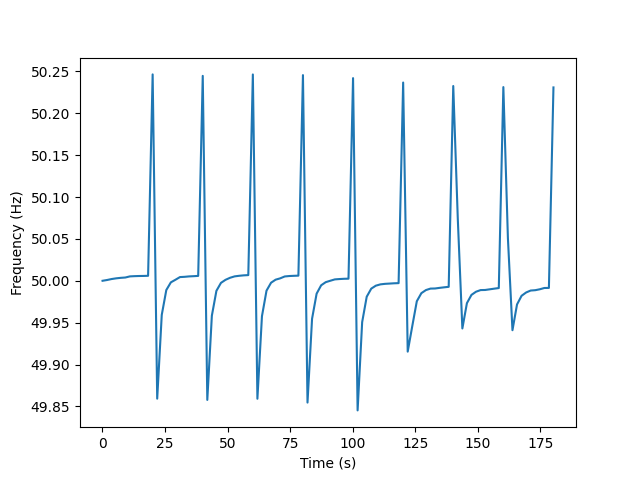}
	\caption{Pulse attack against the sensor-controller link.}
	\centering
	\label{fig:ev_pulse}
\end{figure}

A pulse attack involves occasionally applying a large positive or negative value in a single network packet. Such disruptions can cause a poorly tuned controller to react erratically, and have trouble stabilizing the frequency after this sudden disruption. If the pulses are large and frequent, the system may become unstable. The results displayed in Figure \ref{fig:ev_pulse} indicate that a PID controller with the default tuning can effectively counter pulses of 2 Hz introduced in the frequency measurements at intervals of 20 seconds (every 20th packet). Even though these pulses briefly push the frequency outside normal operational boundaries, the system recovers, preventing significant harm, as per the Swedish transmission system operator~\cite{SVK_pdf} requirements. However, if pulses are large enough with smaller  intervals between them, the system will be unable to recover in time.

\paragraph{Random Attack}\label{ev:sec:random_attack}

\begin{figure}[H] 
	\centering 
	\includegraphics[width=1.1\textwidth, height=9cm]{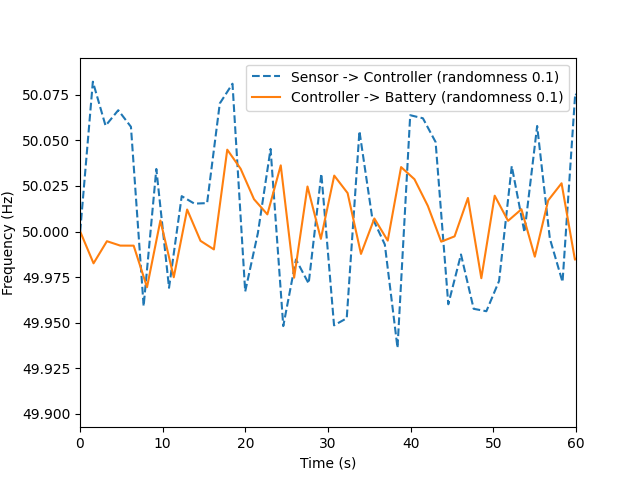}
	\caption{Random attacks against the frequency measurements.}
	\centering
	\label{fig:ev_random}
\end{figure}

A random attack consists of adding a randomness to the frequency measurements or control signals. The randomness is uniformly distributed within some upper and lower amplitude limits. Figure \ref{fig:ev_random} illustrates two examples of such attacks, each with a magnitude of 0.1, targeting the sensor to controller and controller to battery links. These attacks introduce noise, which if large enough, can be difficult for the controller to counteract.

\paragraph{Bias Attack}\label{ev:sec:bias_attack}

\begin{figure}[H] 
	\centering 
	\includegraphics[width=1.1\textwidth]{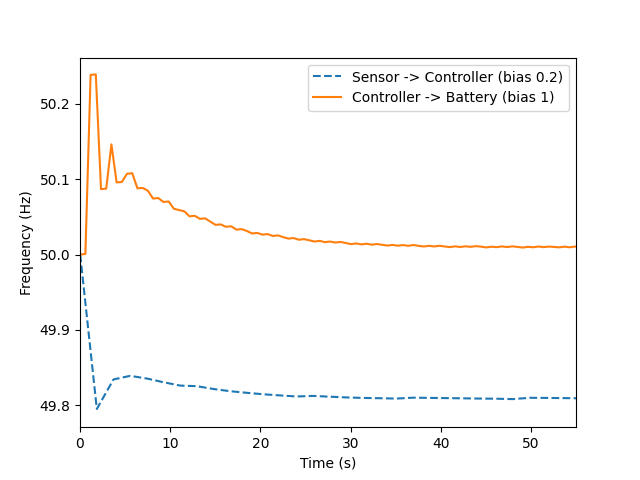}
	\caption{Bias attacks.}
	\centering
	\label{fig:ev_bias}
\end{figure}

In a bias attack, an offset is added to the targeted values. For instance, a constant offset can be added or subtracted to the frequency measurements or the control signals, depending on the sign of the offset. As shown in Figure \ref{fig:ev_bias}, applying the offset on the link between the sensor and controller pushes the frequency to shift in the direction of the offset. For example, an offset of 0.2 stabilizes the system at 49.8 Hz. On the controller-battery link, an offset of 1 (i.e. half of the maximum battery power) briefly pushes the frequency away from 50 Hz since the battery initially gets a requested balancing power of 1MW. Eventually, the controller learns and adapts to the offset by adjusting its control signals to mitigate the offset. In conclusion, a bias attack on the frequency measurements can be more damaging because it misleads the controller into making adjustments based on an incorrect value, unaware of the actual frequency value.

\subsubsection{Replay attack}

\begin{figure}[H] 
	\centering 
	\includegraphics[width=1.1\textwidth]{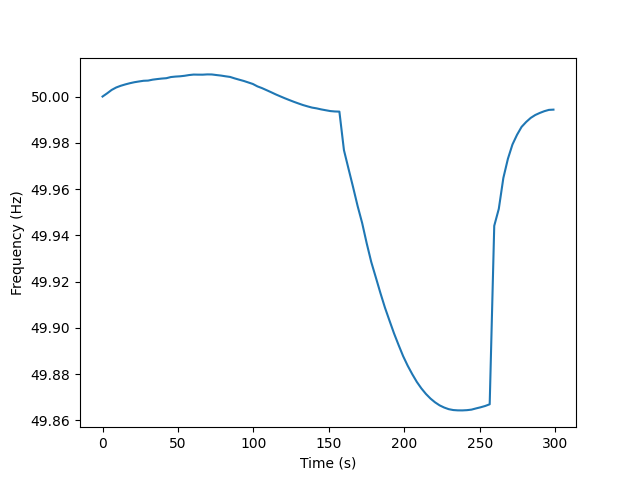}
	\caption{A replay attack against the sensor-controller link.}
	\centering
	\label{fig:ev_replay}
\end{figure}

In a replay attack, a threat actor records network packets on a link and then, after a specified period or a set number of packets, replays the recorded packets. In our context, an attacker could replace the real frequency measurements or control signals with old values. We observed that if the attacker manages to record packets while the consumption is decreasing (such as the first 100s of the consumption dataset for the evaluation as shown in Figure \ref{fig:ev_consumption}), and then replay those values when the consumption instead starts to increase, the controller mistakenly believes the frequency is still at 50 Hz. In reality, however, the frequency starts to deviate. This effect continues until the replay attack is over, at which point the controller realizes that the frequency has deviated and adjusts it back to 50 Hz. These sequence of events are visible in Figure \ref{fig:ev_replay}. In the first 100s, packets are recorded, and during seconds 100-200, the packets are replayed, tricking the controller into believing that the frequency is stable while it actually starts to deviate.

%needs improvement
\subsubsection{Load altering (LA) attack}

A LA attack differs from the previously mentioned attacks in that it requires a threat actor to actually have control over batteries or other power-affecting devices within the electrical grid, in contrast to simply compromising network links. In the simulator, LA attacks are represented by three parameters: interval, offset, and randomness. The interval dictates how often the altering occurs. The other two parameters either add a positive or negative offset or a random value to the consumption data points, similar to how FDI bias and random attacks modify data, before the data is fed into the simulation. 

\begin{figure}[H] 
	\centering 
	\includegraphics[width=1.1\textwidth]{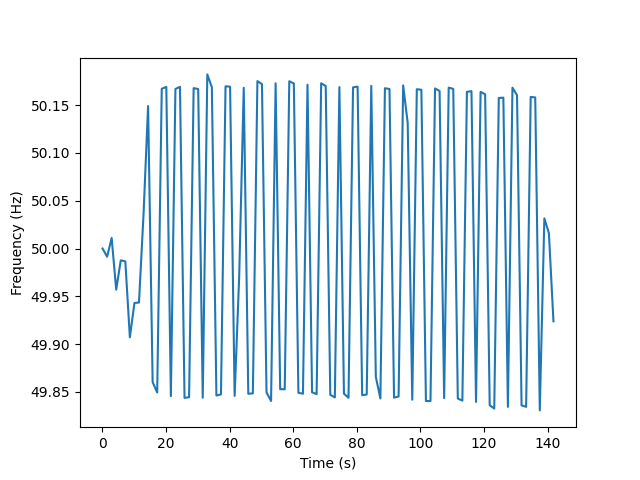}
	\caption{A resourceful LA attack.}
	\centering
	\label{fig:ev_resourceful_load_altering}
\end{figure}

Various types of LA attacks were tested, and the results show that when randomness is as high as half the battery power (1 MW), the system becomes unstable. The impacts of more sophisticated attacks were also investigated. Figure \ref{fig:ev_resourceful_load_altering} shown one example where the frequency was pushed outside the normal operational state (deviating more than 0.1 Hz) by randomness, and then the attacker altered the consumption in the same direction as the battery worked to counteract the deviation (15\%, of the maximum battery power). This caused the system frequency to oscillate rapidly between values outside normal operation. Notably, this attack only required 15\% of the battery power, which makes it highly resourceful given its significant effect on the frequency.

\section{Discussion}
\label{sec:discussion}

%This section discusses the process of evaluating cyberattacks using simulations. It includes reflections on the choice of attacks that were considered along with an assessment of the validity, reliability, and replicability of the resulted obtained. The section also delves into the assumptions and the alternatives considered during the evaluation. Finally, the section outlines potential directions for future work.

The aim of this paper is to provide insights on the structure of cloud-controlled BESSs used to maintain balance in the grid and to help understand the impact of potential cyberattacks on the frequency of the power grid. Based on a literature review, we proposed a reference model for such a system. The reference model outlined key components, interactions, and flows of the system, which enabled systematic and structured identification of the critical assets and potential attack vectors. The identified cyber threats were then evaluated using simulations. 

The proposed RM for a cloud-controlled BESS was the result of an iterative process consisting of literature study and consultations with experts. The RM focuses on the short-term (quasi-real-time) balancing of consumption and production, instead of the so-called load shifting (or peak shaving), where the goal is instead to balance out consumption and production over a day or longer.  Additionally, the RM consists of a control loop that uses frequency to measure the grid's health and not any other control loops such as balancing the voltage or the phase difference between the voltage and current. This scope is reasonable, as TSOs, including SvK, the Swedish TSO, generally only request balancing services for frequency control. 

The identified attack vectors and resulting attacks in the system are a consequence of the scope for RM and the literature review. The most crucial attack vectors identified in the paper are: (i) The internet communication links from FM to CCS and from CCS to BMS (in the control loop), (ii) threat actors seizing control of batteries and activating/deactivating them in a specific pattern, (iii) RA links abused to compromise components in the system and (iv) disturbing cloud dependencies such as DNS and NTP servers and the contracts market. For cyberattacks, the literature highlighted many attacks targeting CCS resulting in the following five attacks: (i) TDS, (ii) DoS, (iii) FDI, (iv) Replay and (v) LA. These are not a comprehensive list of all the possible attack vectors and attacks, but rather a careful selection. Evaluation results showed that attacks that slowly alter sensor readings (such as the ramp attack) could be more dangerous than the ones where there are sudden spikes. This is because the detection systems are typically more sensitive to sudden changes than when values are slightly altered for every reading. Table \ref{tab:conclusion_evaluation} summarizes the effects of different attacks on the simulated frequency. The grid frequency exhibits varying behaviour depending on the type of attack and the specific location within the system that is targeted. The attack vectors are either: sensor to controller (S$\,\to\,$C), controller to battery (C$\,\to\,$B), or the power grid itself in the case of LA attacks.

\begin{table}
	\centering
	\caption{Summary of results.}
	\begin{tabular}{| p{0.2\linewidth} | p{0.2\linewidth} | p{0.5\linewidth} |}
		\hline
		\textbf{Attack} & \textbf{Attack vector} & \textbf{Impact}
		\\ \hline
		Delay & S $\,\to\,$ C + C$\,\to\,$B & Big delays result in frequency oscillations.
		\\ \hline
		DoS & S$\,\to\,$C + C$\,\to\,$B & Packet loss above a certain threshold results in unpredictable system behaviour or frequency oscillations.
		\\ \hline
		Random attack & S$\,\to\,$C & Unpredictable system behaviour or frequency oscillations.
		\\ \hline
		Random attack & C$\,\to\,$B & Unpredictable system behaviour or frequency oscillations, more stable than S$\,\to\,$C.
		\\ \hline
		Bias attack & S$\,\to\,$C & Steady state error.
		\\ \hline
		Scaling attack & S$\,\to\,$C & Steady state error.
		\\ \hline
		Scaling attack & C$\,\to\,$B & Large enough scaling results in frequency oscillations.
		\\ \hline
		Ramp attack & S$\,\to\,$C & Deviation from the nominal frequency.
		\\ \hline
		Ramp attack & C$\,\to\,$B & Large enough ramp value results in steady state error.
		\\ \hline
		Replay attack & S$\,\to\,$C + C$\,\to\,$B & Deviation from the nominal frequency.
		\\ \hline
		Load altering & Grid & Frequency oscillations.
		\\ \hline
	\end{tabular}
	\centering
	\label{tab:conclusion_evaluation}
\end{table}

% We believe we have identified many interesting cases and results with this approach of exploratory and manual testing.

For the evaluation, an exploratory approach was employed, which involved testing different values of the cyberattack parameters with small incremental and decremental adjustments once the system presented some interesting behaviour. It should be noted that the experiments did not run long enough for the batteries to reach a state of charge of 100\% or 0\%. Thus, results regarding the BESS's inability to regulate the frequency when fully charged or empty could not be evaluated. As the data and simulator used in the paper are open-source, the evaluation results can be replicated. Since there is a small inherent randomness built into the simulator, the result may slightly differ, but the general behaviour of each test should remain similar. 

%Real-world power systems use much more complex relationships to determine frequency dynamics, involving inertia, control systems, and generator characteristics.

In the future, it can be beneficial to further extend the RM. One direction could be to add more control loops, such as balancing the voltage or the phase difference between the voltage and current. It is also worth investigating the potential attack vectors on DNS, NTP, contracts, and RA, although they are strictly part of the core control loop. Additionally, the simulator used for evaluation does not consider grid inertia to determine frequency, which may mean that the frequency in the simulation may change more rapidly than in a real-world system. Thus, the simulator can be extended with the notion of inertia for a more realistic assessment. Moreover, multiple sets of consumption data could be tested. For the explored attacks, further testing with different distributions of randomness is another possible future direction. Finally, more attacks can be implemented and tested. It could be worth investigating the usefulness of automating attack testing using a script, that allows for granular testing of each parameter across a wide range of values, while also considering combinations with variations of other parameters. 

\section{Conclusions}
\label{sec:conclusions}

This paper highlights the cybersecurity challenges faced by cloud-controlled battery energy storage systems used for electrical grid frequency regulation. We propose a reference model to provide a structured framework for understanding
the architecture and dynamics of such systems and identify potential attack vectors and cyberattacks. We also evaluate the impact of attacks on the grid frequency and stability using simulations. Evaluations results demonstrate that attacks, particularly those altering data gradually, such as the ramp attack, can severely disrupt grid stability. Our findings underscore the need for robust detection and mitigation strategies for such systems to ensure the resilience and security of the electrical grid in the face of growing cyber threats. 

\backmatter

\section*{Declarations}

\subsection*{Competing interests}
The authors declare that they have no competing interests.

\subsection*{Funding}
The work was conducted as part of the Cybersecurity for Resilient Energy Communities of the Future (CyREC) project, funded by the Swedish innovation agency (reference number 2023-02987), and the FUS21-0033 (ASTECC) project, funded by the Swedish Foundation for Strategic Research (SSF). The funders had no role in the design of the study; in the collection, analyses, or interpretation of data; in the writing of the manuscript; or in the decision to publish the results.

\subsection*{Author contribution}

Writing-original draft, Z.A. and M.A; conceptualisation, F.Ö, J.O, J.D and M.A.; methodology, F.Ö., J.O., M.A.; funding acquisition, M.A. All authors have read and agreed to the published version of the manuscript.

\subsection*{Data availability}
The data and code for the simulator used in the study are openly available in the repository at~\cite{repo}.

\subsection*{Acknowledgements}
The authors would like to thank Ahmad Usman for his support during the course of this work. 

\subsection*{Ethical approval}
Not applicable.

\bibliography{references.bib}% common bib file

%% BioMed_Central_Bib_Style_v1.01

\begin{thebibliography}{38}
% BibTex style file: bmc-mathphys.bst (version 2.1), 2014-07-24
\ifx \bisbn   \undefined \def \bisbn  #1{ISBN #1}\fi
\ifx \binits  \undefined \def \binits#1{#1}\fi
\ifx \bauthor  \undefined \def \bauthor#1{#1}\fi
\ifx \batitle  \undefined \def \batitle#1{#1}\fi
\ifx \bjtitle  \undefined \def \bjtitle#1{#1}\fi
\ifx \bvolume  \undefined \def \bvolume#1{\textbf{#1}}\fi
\ifx \byear  \undefined \def \byear#1{#1}\fi
\ifx \bissue  \undefined \def \bissue#1{#1}\fi
\ifx \bfpage  \undefined \def \bfpage#1{#1}\fi
\ifx \blpage  \undefined \def \blpage #1{#1}\fi
\ifx \burl  \undefined \def \burl#1{\textsf{#1}}\fi
\ifx \doiurl  \undefined \def \doiurl#1{\url{https://doi.org/#1}}\fi
\ifx \betal  \undefined \def \betal{\textit{et al.}}\fi
\ifx \binstitute  \undefined \def \binstitute#1{#1}\fi
\ifx \binstitutionaled  \undefined \def \binstitutionaled#1{#1}\fi
\ifx \bctitle  \undefined \def \bctitle#1{#1}\fi
\ifx \beditor  \undefined \def \beditor#1{#1}\fi
\ifx \bpublisher  \undefined \def \bpublisher#1{#1}\fi
\ifx \bbtitle  \undefined \def \bbtitle#1{#1}\fi
\ifx \bedition  \undefined \def \bedition#1{#1}\fi
\ifx \bseriesno  \undefined \def \bseriesno#1{#1}\fi
\ifx \blocation  \undefined \def \blocation#1{#1}\fi
\ifx \bsertitle  \undefined \def \bsertitle#1{#1}\fi
\ifx \bsnm \undefined \def \bsnm#1{#1}\fi
\ifx \bsuffix \undefined \def \bsuffix#1{#1}\fi
\ifx \bparticle \undefined \def \bparticle#1{#1}\fi
\ifx \barticle \undefined \def \barticle#1{#1}\fi
\bibcommenthead
\ifx \bconfdate \undefined \def \bconfdate #1{#1}\fi
\ifx \botherref \undefined \def \botherref #1{#1}\fi
\ifx \url \undefined \def \url#1{\textsf{#1}}\fi
\ifx \bchapter \undefined \def \bchapter#1{#1}\fi
\ifx \bbook \undefined \def \bbook#1{#1}\fi
\ifx \bcomment \undefined \def \bcomment#1{#1}\fi
\ifx \oauthor \undefined \def \oauthor#1{#1}\fi
\ifx \citeauthoryear \undefined \def \citeauthoryear#1{#1}\fi
\ifx \endbibitem  \undefined \def \endbibitem {}\fi
\ifx \bconflocation  \undefined \def \bconflocation#1{#1}\fi
\ifx \arxivurl  \undefined \def \arxivurl#1{\textsf{#1}}\fi
\csname PreBibitemsHook\endcsname

%%% 1
\bibitem[\protect\citeauthoryear{Rinaldi et~al.}{2022}]{Rinaldi_2022}
\begin{bchapter}
\bauthor{\bsnm{Rinaldi}, \binits{G.}},
\bauthor{\bsnm{Cucuzzella}, \binits{M.}},
\bauthor{\bsnm{Menon}, \binits{P.P.}},
\bauthor{\bsnm{Ferrara}, \binits{A.}},
\bauthor{\bsnm{Edwards}, \binits{C.}}:
\bctitle{Load altering attacks detection, reconstruction and mitigation for
  cyber-security in smart grids with battery energy storage systems}.
In: \bbtitle{2022 European Control Conference (ECC)},
pp. \bfpage{1541}--\blpage{1547}
(\byear{2022}).
\doiurl{10.23919/ECC55457.2022.9838515}
\end{bchapter}
\endbibitem

%%% 2
\bibitem[\protect\citeauthoryear{Oshnoei et~al.}{2020}]{Oshnoei_2020}
\begin{barticle}
\bauthor{\bsnm{Oshnoei}, \binits{A.}},
\bauthor{\bsnm{Kheradmandi}, \binits{M.}},
\bauthor{\bsnm{Muyeen}, \binits{S.M.}}:
\batitle{Robust control scheme for distributed battery energy storage systems
  in load frequency control}.
\bjtitle{IEEE Transactions on Power Systems}
\bvolume{35}(\bissue{6}),
\bfpage{4781}--\blpage{4791}
(\byear{2020})
\doiurl{10.1109/TPWRS.2020.2997950}
\end{barticle}
\endbibitem

%%% 3
\bibitem[\protect\citeauthoryear{Saini et~al.}{2023}]{Saini_2023}
\begin{bchapter}
\bauthor{\bsnm{Saini}, \binits{V.K.}},
\bauthor{\bsnm{Yelisetti}, \binits{S.}},
\bauthor{\bsnm{Kumar}, \binits{R.}},
\bauthor{\bsnm{Al-Sumaiti}, \binits{A.S.}}:
\bctitle{Cloud energy storage management including smart home physical
  parameters}.
In: \bbtitle{2023 IEEE IAS Global Conference on Emerging Technologies
  (GlobConET)},
pp. \bfpage{1}--\blpage{6}
(\byear{2023}).
\doiurl{10.1109/GlobConET56651.2023.10150077}
\end{bchapter}
\endbibitem

%%% 4
\bibitem[\protect\citeauthoryear{Trevizan et~al.}{2022}]{Trevizan_2022}
\begin{barticle}
\bauthor{\bsnm{Trevizan}, \binits{R.D.}},
\bauthor{\bsnm{Obert}, \binits{J.}},
\bauthor{\bsnm{De~Angelis}, \binits{V.}},
\bauthor{\bsnm{Nguyen}, \binits{T.A.}},
\bauthor{\bsnm{Rao}, \binits{V.S.}},
\bauthor{\bsnm{Chalamala}, \binits{B.R.}}:
\batitle{Cyberphysical security of grid battery energy storage systems}.
\bjtitle{IEEE Access}
\bvolume{10},
\bfpage{59675}--\blpage{59722}
(\byear{2022})
\doiurl{10.1109/ACCESS.2022.3178987}
\end{barticle}
\endbibitem

%%% 5
\bibitem[\protect\citeauthoryear{Mohan et~al.}{2020}]{mohan2020comprehensive}
\begin{barticle}
\bauthor{\bsnm{Mohan}, \binits{A.M.}},
\bauthor{\bsnm{Meskin}, \binits{N.}},
\bauthor{\bsnm{Mehrjerdi}, \binits{H.}}:
\batitle{A comprehensive review of the cyber-attacks and cyber-security on load
  frequency control of power systems}.
\bjtitle{Energies}
\bvolume{13}(\bissue{15}),
\bfpage{3860}
(\byear{2020})
\doiurl{10.3390/en13153860}
\end{barticle}
\endbibitem

%%% 6
\bibitem[\protect\citeauthoryear{Ibraheem et~al.}{2023}]{Ibraheem_2023}
\begin{botherref}
\oauthor{\bsnm{Ibraheem}, \binits{M.I.}},
\oauthor{\bsnm{Edrisi}, \binits{M.}},
\oauthor{\bsnm{Alhelou}, \binits{H.H.}},
\oauthor{\bsnm{Gholipour}, \binits{M.}},
\oauthor{\bsnm{Al-Hinai}, \binits{A.}}:
A sophisticated slide mode controller of microgrid system load frequency
  control under false data injection attack and actuator time delay.
IEEE Transactions on Industry Applications,
1--10
(2023)
\doiurl{10.1109/TIA.2023.3316190}
\end{botherref}
\endbibitem

%%% 7
\bibitem[\protect\citeauthoryear{Darup et~al.}{2020}]{cloud2}
\begin{botherref}
\oauthor{\bsnm{Darup}, \binits{M.S.}},
\oauthor{\bsnm{Alexandru}, \binits{A.B.}},
\oauthor{\bsnm{Quevedo}, \binits{D.E.}},
\oauthor{\bsnm{Pappas}, \binits{G.J.}}:
Encrypted control for networked systems - an illustrative introduction and
  current challenges.
CoRR
\textbf{abs/2010.00268}
(2020)
{\href{https://arxiv.org/abs/2010.00268}{{2010.00268}}}
\end{botherref}
\endbibitem

%%% 8
\bibitem[\protect\citeauthoryear{Xu and Zhu}{2015}]{cloud1}
\begin{bchapter}
\bauthor{\bsnm{Xu}, \binits{Z.}},
\bauthor{\bsnm{Zhu}, \binits{Q.}}:
\bctitle{Secure and resilient control design for cloud enabled networked
  control systems}.
In: \beditor{\bsnm{Ray}, \binits{I.}},
\beditor{\bsnm{Thomas}, \binits{R.K.}},
\beditor{\bsnm{C{\'{a}}rdenas}, \binits{A.A.}} (eds.)
\bbtitle{Proceedings of the First {ACM} Workshop on Cyber-Physical
  Systems-Security And/or PrivaCy, {CPS-SPC} 2015, Denver, Colorado, USA,
  October 16, 2015},
pp. \bfpage{31}--\blpage{42}.
\bpublisher{{ACM}}, \blocation{???}
(\byear{2015}).
\doiurl{10.1145/2808705.2808708} .
\burl{https://doi.org/10.1145/2808705.2808708}
\end{bchapter}
\endbibitem

%%% 9
\bibitem[\protect\citeauthoryear{Nguyen et~al.}{2020}]{powergridresiliance}
\begin{barticle}
\bauthor{\bsnm{Nguyen}, \binits{T.}},
\bauthor{\bsnm{Wang}, \binits{S.}},
\bauthor{\bsnm{Alhazmi}, \binits{M.}},
\bauthor{\bsnm{Nazemi}, \binits{M.}},
\bauthor{\bsnm{Estebsari}, \binits{A.}},
\bauthor{\bsnm{Dehghanian}, \binits{P.}}:
\batitle{Electric power grid resilience to cyber adversaries: State of the
  art}.
\bjtitle{IEEE Access}
\bvolume{8},
\bfpage{87592}--\blpage{87608}
(\byear{2020})
\doiurl{10.1109/ACCESS.2020.2993233}
\end{barticle}
\endbibitem

%%% 10
\bibitem[\protect\citeauthoryear{Oscarsson and {\"O}hrstr{\"o}m}{2024}]{thesis}
\begin{botherref}
\oauthor{\bsnm{Oscarsson}, \binits{J.}},
\oauthor{\bsnm{{\"O}hrstr{\"o}m}, \binits{F.}}:
Cyberattack evaluation of cloud-controlled energy storage.
Master's thesis
(2024).
\url{https://urn.kb.se/resolve?urn=urn:nbn:se:liu:diva-205693}
\end{botherref}
\endbibitem

%%% 11
\bibitem[\protect\citeauthoryear{Baumgart et~al.}{2019}]{Baumgart_2019}
\begin{bchapter}
\bauthor{\bsnm{Baumgart}, \binits{I.}},
\bauthor{\bsnm{Borsig}, \binits{M.}},
\bauthor{\bsnm{Goerke}, \binits{N.}},
\bauthor{\bsnm{Hackenjos}, \binits{T.}},
\bauthor{\bsnm{Rill}, \binits{J.}},
\bauthor{\bsnm{Wehmer}, \binits{M.}}:
\bctitle{Who controls your energy? on the (in)security of residential battery
  energy storage systems}.
In: \bbtitle{2019 IEEE International Conference on Communications, Control, and
  Computing Technologies for Smart Grids (SmartGridComm)},
pp. \bfpage{1}--\blpage{6}
(\byear{2019}).
\doiurl{10.1109/SmartGridComm.2019.8909749}
\end{bchapter}
\endbibitem

%%% 12
\bibitem[\protect\citeauthoryear{Naseri et~al.}{2023}]{Naseri_2023}
\begin{barticle}
\bauthor{\bsnm{Naseri}, \binits{F.}},
\bauthor{\bsnm{Kazemi}, \binits{Z.}},
\bauthor{\bsnm{Larsen}, \binits{P.G.}},
\bauthor{\bsnm{Arefi}, \binits{M.M.}},
\bauthor{\bsnm{Schaltz}, \binits{E.}}:
\batitle{Cyber-physical cloud battery management systems: Review of security
  aspects}.
\bjtitle{Batteries}
(\byear{2023})
\doiurl{10.3390/batteries9070382}
\end{barticle}
\endbibitem

%%% 13
\bibitem[\protect\citeauthoryear{Kharlamova et~al.}{2020}]{Kharlamova_2020}
\begin{bchapter}
\bauthor{\bsnm{Kharlamova}, \binits{N.}},
\bauthor{\bsnm{Hashemi}, \binits{S.}},
\bauthor{\bsnm{Træholt}, \binits{C.}}:
\bctitle{The cyber security of battery energy storage systems and adoption of
  data-driven methods}.
In: \bbtitle{2020 IEEE Third International Conference on Artificial
  Intelligence and Knowledge Engineering (AIKE)},
pp. \bfpage{188}--\blpage{192}
(\byear{2020}).
\doiurl{10.1109/AIKE48582.2020.00037}
\end{bchapter}
\endbibitem

%%% 14
\bibitem[\protect\citeauthoryear{S{\'a}nchez
  et~al.}{2019}]{sanchez2019bibliographical}
\begin{barticle}
\bauthor{\bsnm{S{\'a}nchez}, \binits{H.S.}},
\bauthor{\bsnm{Rotondo}, \binits{D.}},
\bauthor{\bsnm{Escobet}, \binits{T.}},
\bauthor{\bsnm{Puig}, \binits{V.}},
\bauthor{\bsnm{Quevedo}, \binits{J.}}:
\batitle{Bibliographical review on cyber attacks from a control oriented
  perspective}.
\bjtitle{Annual Reviews in Control}
\bvolume{48},
\bfpage{103}--\blpage{128}
(\byear{2019})
\doiurl{10.1016/j.arcontrol.2019.08.002}
\end{barticle}
\endbibitem

%%% 15
\bibitem[\protect\citeauthoryear{Ghiasi et~al.}{2023}]{mohammad2023}
\begin{barticle}
\bauthor{\bsnm{Ghiasi}, \binits{M.}},
\bauthor{\bsnm{Niknam}, \binits{T.}},
\bauthor{\bsnm{Wang}, \binits{Z.}},
\bauthor{\bsnm{Mehrandezh}, \binits{M.}},
\bauthor{\bsnm{Dehghani}, \binits{M.}},
\bauthor{\bsnm{Ghadimi}, \binits{N.}}:
\batitle{A comprehensive review of cyber-attacks and defense mechanisms for
  improving security in smart grid energy systems: Past, present and future}.
\bjtitle{Electric Power Systems Research}
\bvolume{215},
\bfpage{108975}
(\byear{2023})
\doiurl{10.1016/j.epsr.2022.108975}
\end{barticle}
\endbibitem

%%% 16
\bibitem[\protect\citeauthoryear{Mahmoud et~al.}{2019}]{mahmoud2019modeling}
\begin{barticle}
\bauthor{\bsnm{Mahmoud}, \binits{M.S.}},
\bauthor{\bsnm{Hamdan}, \binits{M.M.}},
\bauthor{\bsnm{Baroudi}, \binits{U.A.}}:
\batitle{Modeling and control of cyber-physical systems subject to cyber
  attacks: A survey of recent advances and challenges}.
\bjtitle{Neurocomputing}
\bvolume{338},
\bfpage{101}--\blpage{115}
(\byear{2019})
\doiurl{10.1016/j.neucom.2019.01.099}
\end{barticle}
\endbibitem

%%% 17
\bibitem[\protect\citeauthoryear{Wu et~al.}{2019}]{Wu_2019}
\begin{bchapter}
\bauthor{\bsnm{Wu}, \binits{Y.}},
\bauthor{\bsnm{Weng}, \binits{J.}},
\bauthor{\bsnm{Qiu}, \binits{B.}},
\bauthor{\bsnm{Wei}, \binits{Z.}},
\bauthor{\bsnm{Qian}, \binits{F.}},
\bauthor{\bsnm{Deng}, \binits{R.H.}}:
\bctitle{Random delay attack and its applications on load frequency control of
  power systems}.
In: \bbtitle{2019 IEEE Conference on Dependable and Secure Computing (DSC)},
pp. \bfpage{1}--\blpage{8}
(\byear{2019}).
\doiurl{10.1109/DSC47296.2019.8937611}
\end{bchapter}
\endbibitem

%%% 18
\bibitem[\protect\citeauthoryear{Chauhan and Shiaeles}{2023}]{Chauhan_2023}
\begin{barticle}
\bauthor{\bsnm{Chauhan}, \binits{M.}},
\bauthor{\bsnm{Shiaeles}, \binits{S.}}:
\batitle{An analysis of cloud security frameworks, problems and proposed
  solutions}.
\bjtitle{Network}
\bvolume{3}(\bissue{3}),
\bfpage{422}--\blpage{450}
(\byear{2023})
\doiurl{10.3390/network3030018}
\end{barticle}
\endbibitem

%%% 19
\bibitem[\protect\citeauthoryear{Akbarian et~al.}{2023}]{Akbarian_2023}
\begin{bchapter}
\bauthor{\bsnm{Akbarian}, \binits{F.}},
\bauthor{\bsnm{Tärneberg}, \binits{W.}},
\bauthor{\bsnm{Fitzgerald}, \binits{E.}},
\bauthor{\bsnm{Kihl}, \binits{M.}}:
\bctitle{Detecting and mitigating actuator attacks on cloud control systems
  through digital twins}.
In: \bbtitle{2023 International Conference on Software, Telecommunications and
  Computer Networks (SoftCOM)},
pp. \bfpage{1}--\blpage{6}
(\byear{2023}).
\doiurl{10.23919/SoftCOM58365.2023.10271648}
\end{bchapter}
\endbibitem

%%% 20
\bibitem[\protect\citeauthoryear{Ghiasi et~al.}{2021}]{Ghiasi_2021}
\begin{barticle}
\bauthor{\bsnm{Ghiasi}, \binits{M.}},
\bauthor{\bsnm{Dehghani}, \binits{M.}},
\bauthor{\bsnm{Niknam}, \binits{T.}},
\bauthor{\bsnm{Kavousi-Fard}, \binits{A.}},
\bauthor{\bsnm{Siano}, \binits{P.}},
\bauthor{\bsnm{Alhelou}, \binits{H.H.}}:
\batitle{Cyber-attack detection and cyber-security enhancement in smart
  dc-microgrid based on blockchain technology and hilbert huang transform}.
\bjtitle{IEEE Access}
\bvolume{9},
\bfpage{29429}--\blpage{29440}
(\byear{2021})
\doiurl{10.1109/ACCESS.2021.3059042}
\end{barticle}
\endbibitem

%%% 21
\bibitem[\protect\citeauthoryear{Gumrukcu et~al.}{2022}]{Gumrukcu_2022}
\begin{bchapter}
\bauthor{\bsnm{Gumrukcu}, \binits{E.}},
\bauthor{\bsnm{Arsalan}, \binits{A.}},
\bauthor{\bsnm{Muriithi}, \binits{G.}},
\bauthor{\bsnm{Joglekar}, \binits{C.}},
\bauthor{\bsnm{Aboulebdeh}, \binits{A.}},
\bauthor{\bsnm{Alparslan~Zehir}, \binits{M.}},
\bauthor{\bsnm{Papari}, \binits{B.}},
\bauthor{\bsnm{Monti}, \binits{A.}}:
\bctitle{Impact of cyber-attacks on ev charging coordination: The case of
  single point of failure}.
In: \bbtitle{2022 4th Global Power, Energy and Communication Conference
  (GPECOM)},
pp. \bfpage{506}--\blpage{511}
(\byear{2022}).
\doiurl{10.1109/GPECOM55404.2022.9815727}
\end{bchapter}
\endbibitem

%%% 22
\bibitem[\protect\citeauthoryear{Le et~al.}{2020}]{GridAttackSim}
\begin{botherref}
\oauthor{\bsnm{Le}, \binits{T.D.}},
\oauthor{\bsnm{Anwar}, \binits{A.}},
\oauthor{\bsnm{Loke}, \binits{S.W.}},
\oauthor{\bsnm{Beuran}, \binits{R.}},
\oauthor{\bsnm{Tan}, \binits{Y.}}:
Gridattacksim: A cyber attack simulation framework for smart grids.
Electronics
\textbf{9}(8)
(2020)
\doiurl{10.3390/electronics9081218}
\end{botherref}
\endbibitem

%%% 23
\bibitem[\protect\citeauthoryear{Oest et~al.}{2023}]{cosima}
\begin{barticle}
\bauthor{\bsnm{Oest}, \binits{F.}},
\bauthor{\bsnm{Frost}, \binits{E.}},
\bauthor{\bsnm{Radtke}, \binits{M.}},
\bauthor{\bsnm{Lehnhoff}, \binits{S.}}:
\batitle{Coupling omnet++ and mosaik for integrated co-simulation of
  ict-reliant smart grids}.
\bjtitle{SIGENERGY Energy Inform. Rev.}
\bvolume{3}(\bissue{1}),
\bfpage{14}--\blpage{25}
(\byear{2023})
\doiurl{10.1145/3607120.3607123}
\end{barticle}
\endbibitem

%%% 24
\bibitem[\protect\citeauthoryear{Kumar and Bhongade}{2016}]{Kumar_2016}
\begin{bchapter}
\bauthor{\bsnm{Kumar}, \binits{B.}},
\bauthor{\bsnm{Bhongade}, \binits{S.}}:
\bctitle{Load disturbance rejection based pid controller for frequency
  regulation of a microgrid}.
In: \bbtitle{2016 Biennial International Conference on Power and Energy
  Systems: Towards Sustainable Energy (PESTSE)},
pp. \bfpage{1}--\blpage{6}
(\byear{2016}).
\doiurl{10.1109/PESTSE.2016.7516459}
\end{bchapter}
\endbibitem

%%% 25
\bibitem[\protect\citeauthoryear{}{}]{Cyrec}
\begin{botherref}
Cybersecurity for Resilient Energy Communities of the Future.
Accessed: 2025-01-01.
\url{https://www.vinnova.se/en/p/cybersecurity-for-resilient-energy-communities-of-the-future/}
\end{botherref}
\endbibitem

%%% 26
\bibitem[\protect\citeauthoryear{Fette and Melnikov}{2011}]{Fette_2011}
\begin{botherref}
\oauthor{\bsnm{Fette}, \binits{I.}},
\oauthor{\bsnm{Melnikov}, \binits{A.}}:
The websocket protocol.
Technical report
(2011).
\doiurl{10.17487/RFC6455}
\end{botherref}
\endbibitem

%%% 27
\bibitem[\protect\citeauthoryear{Behera et~al.}{2022}]{Behera_2022}
\begin{barticle}
\bauthor{\bsnm{Behera}, \binits{B.B.}},
\bauthor{\bsnm{Mohanty}, \binits{R.K.}},
\bauthor{\bsnm{Pattanayak}, \binits{B.K.}}:
\batitle{A deep fusion model for automated industrial iot cyber attack
  detection and mitigation}.
\bjtitle{IJEER}
\bvolume{10}(\bissue{3}),
\bfpage{604}--\blpage{613}
(\byear{2022})
\doiurl{10.37391/IJEER.100332}
\end{barticle}
\endbibitem

%%% 28
\bibitem[\protect\citeauthoryear{Ganesan et~al.}{2017}]{Ganesan_2017}
\begin{bbook}
\bauthor{\bsnm{Ganesan}, \binits{R.}},
\bauthor{\bsnm{Shah}, \binits{A.}},
\bauthor{\bsnm{Jajodia}, \binits{S.}},
\bauthor{\bsnm{Cam}, \binits{H.}}:
\bbtitle{A Novel Metric for Measuring Operational Effectiveness of a
  Cybersecurity Operations Center},
pp. \bfpage{177}--\blpage{207}.
\bpublisher{Springer},
\blocation{Cham}
(\byear{2017}).
\doiurl{10.1007/978-3-319-66505-4_8}
\end{bbook}
\endbibitem

%%% 29
\bibitem[\protect\citeauthoryear{{de Souza} et~al.}{2020}]{Maestro_2020}
\begin{bchapter}
\bauthor{\bsnm{{de Souza}}, \binits{E.}},
\bauthor{\bsnm{{Ardakanian}}, \binits{O.}},
\bauthor{\bsnm{{Nikolaidis}}, \binits{I.}}:
\bctitle{A co-simulation platform for evaluating cyber security and control
  applications in the smart grid}.
In: \bbtitle{ICC 2020 - 2020 IEEE International Conference on Communications
  (ICC)},
pp. \bfpage{1}--\blpage{7}
(\byear{2020}).
\doiurl{10.1109/ICC40277.2020.9149212}
\end{bchapter}
\endbibitem

%%% 30
\bibitem[\protect\citeauthoryear{Hardy et~al.}{2024}]{HELICS}
\begin{barticle}
\bauthor{\bsnm{Hardy}, \binits{T.D.}},
\bauthor{\bsnm{Palmintier}, \binits{B.}},
\bauthor{\bsnm{Top}, \binits{P.L.}},
\bauthor{\bsnm{Krishnamurthy}, \binits{D.}},
\bauthor{\bsnm{Fuller}, \binits{J.C.}}:
\batitle{Helics: A co-simulation framework for scalable multi-domain modeling
  and analysis}.
\bjtitle{IEEE Access}
\bvolume{12},
\bfpage{24325}--\blpage{24347}
(\byear{2024})
\doiurl{10.1109/ACCESS.2024.3363615}
\end{barticle}
\endbibitem

%%% 31
\bibitem[\protect\citeauthoryear{Oscarsson and {\"O}hrstr{\"o}m}{2024}]{repo}
\begin{botherref}
\oauthor{\bsnm{Oscarsson}, \binits{J.}},
\oauthor{\bsnm{{\"O}hrstr{\"o}m}, \binits{F.}}:
The Simulator.
\url{https://gitlab.liu.se/joaos226/cyberattack-evaluation-of-cloud-controlled-energy-storage}
Accessed Accessed: 2024-10-10
\end{botherref}
\endbibitem

%%% 32
\bibitem[\protect\citeauthoryear{}{}]{SVK_frekvensstabilitet}
\begin{botherref}
Frekvensstabilitet - Svenska Kraftnät.
Accessed: 2025-01-01
\end{botherref}
\endbibitem

%%% 33
\bibitem[\protect\citeauthoryear{Modig et~al.}{2022}]{modig2022overview}
\begin{botherref}
\oauthor{\bsnm{Modig}, \binits{N.}},
\oauthor{\bsnm{Eriksson}, \binits{R.}},
\oauthor{\bsnm{Ruokolainen}, \binits{P.}},
\oauthor{\bsnm{{\O}deg{\aa}rd}, \binits{J.N.}},
\oauthor{\bsnm{Weizenegger}, \binits{S.}},
\oauthor{\bsnm{Fechtenburg}, \binits{T.D.}}:
Overview of frequency control in the nordic power system.
Nordic Analysis Group
(2022)
\end{botherref}
\endbibitem

%%% 34
\bibitem[\protect\citeauthoryear{}{}]{SvK_kontrollrummet}
\begin{botherref}
Kontrollrummet - Svenska Kraftnät.
Accessed: 2025-01-01.
\url{https://www.svk.se/om-kraftsystemet/kontrollrummet/}
\end{botherref}
\endbibitem

%%% 35
\bibitem[\protect\citeauthoryear{}{}]{Vattenfall_elkonsumption}
\begin{botherref}
Normal Elförbrukning För Villa \& Lägenhet.
Accessed: 2025-01-01.
\url{https://www.vattenfall.se/fokus/tips-rad/vad-ar-normal-elforbrukning/}
\end{botherref}
\endbibitem

%%% 36
\bibitem[\protect\citeauthoryear{}{}]{Ny_teknik}
\begin{botherref}
Unik Kartläggning: Batteriparker ökar Enormt – Risk För överetablering.
Accessed: 2025-01-01.
\url{https://www.nyteknik.se/energi/unik-kartlaggning-batteriparker-okar-enormt-risk-for-overetablering/4230110}
\end{botherref}
\endbibitem

%%% 37
\bibitem[\protect\citeauthoryear{}{}]{Pixii}
\begin{botherref}
Pixii Home.
Accessed: 2025-01-01.
\url{https://www.pixii.com/wp-content/uploads/2024/05/2024-05-26_Ver.2.3_Pixii_Home.pdf}
\end{botherref}
\endbibitem

%%% 38
\bibitem[\protect\citeauthoryear{}{}]{SVK_pdf}
\begin{botherref}
Drifttillstånden - Svenska Kraftnät.
Accessed: 2025-01-01.
\url{https://www.svk.se/contentassets/7dfa08444dab4d3aba1adbbc38456c2b/systemdrifttillstanden.pdf}
\end{botherref}
\endbibitem

\end{thebibliography}
%% if required, the content of .bbl file can be included here once bbl is generated
%%\input sn-article.bbl

\end{document}